\documentclass[aps,prl,reprint,groupedaddress]{revtex4-1}

\usepackage{graphicx}
\usepackage{color}
\usepackage{amsmath}
\usepackage[T1]{fontenc}
\usepackage[english]{babel}
\usepackage{hyperref}
\renewcommand{\frac}{\dfrac}

\begin{document}

%\preprint{}

\title{Thermally activated vapor bubble nucleation: \\ the Landau--Lifshitz/Van der Waals approach}
% \title{Bubble nucleation captured by Fluctuating Hydrodynamics: \\a two-phase continuum model accounting for thermal fluctuations}

\author{Mirko Gallo}
\author{Francesco Magaletti}%
\author{Carlo Massimo Casciola}
\email{carlomassimo.casciola@uniroma1.it}
\homepage{http://flumacs.site.uniroma1.it}
\affiliation{ 
Department of Mechanical and Aerospace Engineering, Sapienza Universit\`a di Roma, Rome, Italy%\\This line break forced with \textbackslash\textbackslash
}%

\date{\today}

\begin{abstract}
Vapor bubbles are formed in liquids by two mechanisms: evaporation (temperature above the boiling threshold) and  cavitation
(pressure below the vapor pressure).  The liquid resists in these metastable (overheating and tensile, respectively) states for a long time since
bubble nucleation is  an activated process that needs to surmount the free energy barrier separating the liquid and the vapor states.
The bubble nucleation rate is difficult to assess and, typically,  only for extremely small systems treated at atomistic level of detail. In this work a powerful approach, based on a continuum diffuse interface modeling of the two-phase fluid embedded with thermal fluctuations (Fluctuating Hydrodynamics) is exploited to study the nucleation process in homogeneous conditions, evaluating the bubble nucleation rates and following the long term dynamics of the metastable system, up to the bubble coalescence and expansion stages. In comparison with more classical approaches, this methodology allows on the one hand to deal with much larger systems observed for a much longer  times than possible with even the most advanced atomistic models.  On the other it extends continuum formulations to thermally activated processes, impossible to deal with in a purely determinist setting.
\end{abstract}
\pacs{}
%\keywords{}

\keywords{bubble, nucleation, cavitation, diffuse interface, fluctuating hydrodynamics, thermal fluctuations}
\maketitle

Thermal fluctuations play a dominant role in the dynamics of fluid systems below the micrometer scale. 
Their effects are significant in, e.g., the smallest micro-fluidic devices \cite{bocquet1, bocquet2} or in biological systems such as lipid membranes \cite{naji2009hybrid},  for Brownian engines and in  artificial molecular motors \cite{peskin1993cellular}. 
They are crucial for thermally activated processes such as nucleation, the precursor of the phase change in metastable systems.
Nucleation is directly connected to the phenomenon of bubble cavitation \cite{brennen2013cavitation} and of freezing rain \cite{cao2009anti}, to cite a few.
There, thermal fluctuations allow to overcome the energy barriers for phase transitions \cite{jones1999bubble, kashchiev2003review, 
lohse2016homogeneous}. Depending on the thermodynamic conditions, the nucleation time may be exceedingly long, the so-called \lq\lq rare-event\rq\rq~issue.
Classical nucleation theory (CNT) \cite{blander1975bubble} provides the basic understanding of the phenomenon which is
nowadays addressed through more sophisticated models like density functional theory (DFT) \cite{oxtoby1988nonclassical, lutsko2011density} or by means of molecular dynamics (MD)  simulations \cite{diemand2014direct}. 
These approaches need to be coupled to specialized techniques for rare events, like the  string method \cite{weinan2002string}, the forward flux sampling \cite{allen2009forward} and the transition path sampling \cite{dellago}, to reliably evaluate the nucleation barrier and determine the transition path  \cite{giacomello2012cassie}. For many real systems they are often computationally too expensive and therefore limited to very small domains.

Here we adopt a mesoscopic continuum approach, embedding stochastic fluctuations, for the numerical simulation of thermally activated bubble nucleation.
Since the pioneering work of Landau and Lifshitz (1958, 1959) \cite{landau1980statistical} several works contributed to the growing field of  \lq\lq Fluctuating Hydrodynamics\rq\rq\, (FH) \cite{fox1970contributions}. More recently the theoretical effort has been followed by a flourishing  of highly specialized numerical methods for the treatment of the stochastic contributions \cite{donev2010accuracy, delong2013temporal, balboa2012staggered, donev2014low}.
The present model is based on a diffuse interface \cite{van1979thermodynamic} description of the two--phase vapor--liquid system \cite{magaletti2015shock} similar to the one recently exploited
% Mettere nella lettera all'editor
%To the best of our knowledge, this is the first work where such an approach is used to investigate the bubble nucleation process. 
%
by Chaudhri {\em et al.} \cite{chaudhri2014modeling}  to address the spinodal decomposition.
The thermodynamic range of applicability of this approach is subjected to some restrictions: 
i) at the very first stage of nucleation the vapor nucleii, smaller than the critical size, need to be numerically resolved; analogously, ii) the thin liquid-vapor interface needs to be captured for the correct evaluation of the capillary stresses; iii) fluctuating hydrodynamics
predicts that the fluctuation intensity grows with the inverse cell volume, $\Delta V$, leading to intense fluctuations, contrary to the assumption of weak
noise needed to derive the model  ($\sqrt{\langle \delta f^2 \rangle}/\langle f \rangle \ll 1$). 
Notwithstanding these restrictions, where it can be applied, this mesoscale approach offers a good level of accuracy 
(as will be shown when discussing the results) at a very cheap computational cost compared to other techniques.
The typical size of the system we simulate on a small computational cluster ($200 \times 200 \times 200 \, {\rm nm^3}$, corresponding to a system
of order $10^8$ atomistic particles) is comparable with one of the largest MD simulations \cite{angelil2014bubble} on a tier-0 machine.
Moreover the simulated time is here $T_{\max} \sim {\rm \mu s}$ to be compared with the MD $T_{max} \sim {\rm ns}$. The enormous difference between the  two time extensions allows us to address the simultaneous nucleation of several vapor bubbles, their expansion, coalescence and, at variance with most of the available methods dealing with quasi-static conditions, the resulting excitation of the macroscopic velocity field.

%%%%%%%%%%%%%%%%%%%%%%%%%%%%
The diffuse interface modeling adopted here has a strict relationship with more fundamental atomistic approaches, since it is based on a suitable approximation of the free energy functional derived in DFT \cite{lutsko2011density}.
It dates back to the famous Van der Waals  square gradient approximation of the Helmholtz free energy functional
 $F[\rho,\theta]=\int_{V}\,dV (f_0\left(\rho,\theta\right)+ 1/2\lambda\boldsymbol{\nabla}\rho\cdot\boldsymbol{\nabla}\rho)$,
%
%\begin{equation}
%\label{eq:FreeEnFunctional}
%F\left[\rho,\theta\right]=\int_{V}\,dV \left(f_0\left(\rho,\theta\right)+\frac12\lambda\boldsymbol{\nabla}\rho\cdot\boldsymbol{\nabla}\rho \right) \ , 
%\end{equation}
%
where $f_0$ is the classical bulk free energy, expressed as a function of density $\rho$ and temperature $\theta$.
$\lambda$ is the capillarity coefficient that controls the (equilibrium) surface tension $\gamma$ and interface thickness, e.g. 
$\gamma(\theta) = \int_{\rho_v^{sat}}^{\rho_l^{sat}} \sqrt{2 \lambda \left[\omega_0(\rho) - \omega_0(\rho_v^{sat}) \right]} d \rho$ with $\omega_0(\rho,\theta)$ the bulk grand potential per unit volume and the superscript $sat$ denoting the (temperature dependent) saturation conditions, see Supplemental Material \cite{suppl_mat}.
The mesoscopic fields obey mass, momentum and energy conservation, with the addition of stochastic contributions (Lifshitz-Landau-Navier-Stokes equations with capillarity):
% $\partial_t\rho+ \boldsymbol{\nabla}\cdot( \rho\mathbf{u})=0$, $\partial_t\rho\mathbf{u}+ \boldsymbol{\nabla}\cdot( \rho\mathbf{u}\otimes\mathbf{u}) =-\boldsymbol{\nabla} p +\boldsymbol{\nabla}\cdot\boldsymbol{\Sigma}  +\boldsymbol{\nabla}\cdot\boldsymbol{\delta\Sigma}$, 
%  $\partial_t E + \boldsymbol{\nabla}\cdot ( \mathbf{u} E ) = \boldsymbol{\nabla}\cdot( - p\mathbf{u} +\mathbf{u}\cdot\boldsymbol{\Sigma} -\mathbf{q}) + \boldsymbol{\nabla}\cdot(  \mathbf{u}\cdot\boldsymbol{\delta\Sigma} -\boldsymbol{\delta q})$, 
%
 \begin{eqnarray} \label{eq:LLNS3}
  \dfrac{\partial\rho}{\partial t}+ \boldsymbol{\nabla}\cdot\left( \rho\mathbf{u}\right) &=&0 \,,  \\
  \nonumber
  \dfrac{\partial\rho\mathbf{u}}{\partial t}+ \boldsymbol{\nabla}\cdot\left( \rho\mathbf{u}\otimes\mathbf{u}\right) &=&-\boldsymbol{\nabla} p +\boldsymbol{\nabla}\cdot\boldsymbol{\Sigma}  +\boldsymbol{\nabla}\cdot\boldsymbol{\delta\Sigma}  \,,\\
  \nonumber
  \dfrac{\partial E}{\partial t}+ \boldsymbol{\nabla}\cdot\left( \mathbf{u} E\right) &=&\boldsymbol{\nabla}\cdot\left( - p\mathbf{u} +\mathbf{u}\cdot\boldsymbol{\Sigma} -\mathbf{q} \right) + \\
  \nonumber
  &&+  \boldsymbol{\nabla}\cdot\left(  \mathbf{u}\cdot\boldsymbol{\delta\Sigma} -\boldsymbol{\delta q} \right) \, ,
  \end{eqnarray} 
where $\mathbf{u}$ is the fluid velocity, $p=\rho^2 \partial (f_0/\rho)/\partial \rho$ is the pressure, $E$ is the total energy density, 
$E = {\cal U} + 1/2\rho\vert\mathbf{u}\vert^2 + 1/2 \lambda \vert\boldsymbol{\nabla}\rho\vert^2$, 
with ${\cal U}$ the internal energy density. In the momentum and energy equations, 
$\boldsymbol \Sigma$ and $\mathbf q$ are the classical deterministic stress tensor and energy flux, respectively, and the terms with the prefix 
 $\delta$ are the stochastic parts, required to satisfy a suitable \emph{fluctuation-dissipation balance} (FDB).
The deterministic stress tensor $\boldsymbol{\Sigma} $ and the energy flux $ \boldsymbol{q} $ follow by standard non-equilibrium thermodynamic  considerations \cite{magaletti2016shock} as:
$\boldsymbol{{\Sigma}} = ( \lambda /2 \vert\boldsymbol{\nabla}\rho\vert^2 + \rho\boldsymbol{\nabla}\cdot( \lambda\boldsymbol{\nabla}\rho))\boldsymbol{I} - \lambda\boldsymbol{\nabla}\rho\otimes\boldsymbol{\nabla}\rho +\mu[(\boldsymbol{\nabla}\mathbf{u} + \boldsymbol{\nabla}\mathbf{u}^T) - 2/3\boldsymbol{\nabla}\cdot\mathbf{u}\boldsymbol{I}] $, $\boldsymbol{q} = \lambda \rho \boldsymbol{\nabla}\rho \boldsymbol{\nabla}\cdot \mathbf{u} - k \boldsymbol{\nabla}\theta$, 
%
%\begin{eqnarray}\label{eq:stress_tens}
%\nonumber
%\boldsymbol{{\Sigma}} &=& \left( \frac{\lambda}{2}\vert\boldsymbol{\nabla}\rho\vert^2 + \rho\boldsymbol{\nabla}\cdot\left(\lambda\boldsymbol{\nabla}\rho\right)\right)\boldsymbol{I} - \lambda\boldsymbol{\nabla}\rho\otimes\boldsymbol{\nabla}\rho+ \\
%&&+\mu\left[\left(\boldsymbol{\nabla}\mathbf{u} + \boldsymbol{\nabla}\mathbf{u}^T\right) - \frac{2}{3}\boldsymbol{\nabla}\cdot\mathbf{u}\boldsymbol{I}\right] ,
%\end{eqnarray}
%
%\begin{equation}\label{eq:en_flux}
%\boldsymbol{q} = \lambda \rho \boldsymbol{\nabla}\rho \boldsymbol{\nabla}\cdot \mathbf{u} - k \boldsymbol{\nabla}\theta, 
%\end{equation}
  %
with $\mu$ and $k$ the dynamic viscosity and the thermal conductivity, respectively.
Enforcing the FDB, the covariance of the stochastic fluxes follows as, $\langle \boldsymbol{\delta\Sigma}(\hat{x},\hat{t})\otimes \boldsymbol{\delta\Sigma}^\dagger(\tilde{x},\tilde{t}) \rangle = \mathbf{Q^\Sigma}\delta(\hat{x}-\tilde{x}) \delta(\hat{t}-\tilde{t})$, and $\langle \boldsymbol{\delta q}(\hat{x},\hat{t}) \otimes\boldsymbol{\delta q}^\dagger(\tilde{x},\tilde{t})\rangle= \mathbf{Q^q}\delta(\hat{x}-\tilde{x}) \delta(\hat{t}-\tilde{t})$, 
  %
  %\begin{equation} \label{eq:visc} 	
  %\left\langle \boldsymbol{\delta\Sigma}(\hat{x},\hat{t})\otimes \boldsymbol{\delta\Sigma}^\dagger(\tilde{x},\tilde{t})\right\rangle= \mathbf{Q^\Sigma}\delta(\hat{x}-\tilde{x}) \delta(\hat{t}-\tilde{t}),
  %\end{equation} 
  %
  %\begin{equation} \label{eq:heat}
  %\quad  \left\langle \boldsymbol{\delta q}(\hat{x},\hat{t}) \otimes\boldsymbol{\delta q}^\dagger(\tilde{x},\tilde{t})\right\rangle= \mathbf{Q^q}\delta(\hat{x}-\tilde{x}) \delta(\hat{t}-\tilde{t}),
  %\end{equation} 
  %
  where  
% \begin{equation}
$\mathbf{Q^\Sigma}_{\alpha\beta\nu\eta}=2{\rm k}_B\theta\mu\left( \delta_{\alpha\nu}\delta_{\beta\eta}+\delta_{\alpha\eta}\delta_{\beta\nu}-{2}/{3}\delta_{\alpha\beta}\delta_{\nu\eta}\right)$ and  
% \,, \end{equation}
%
%\begin{equation}
  $ \mathbf{Q^q}_{\alpha\beta}=2{\rm k}_B\theta^2 k\delta_{\alpha\beta}$, with $k_B$ the Boltzmann constant (see the Supplemental Material \cite{suppl_mat} for details).
% \end{equation}   
% 
Thanks to the Curie-Prigogine principle \cite{de2013non},  the cross-correlation between different tensor rank fluxes vanishes,
 \emph{i.e.} ( $\left\langle \delta\mathbf{q}^\dagger(\tilde{x},\tilde{t})\otimes\delta\mathbf{\Sigma}(\hat{x},\hat{t})\right\rangle=0$). 
 %Thus in the theory of fluctuating hydrodynamics, the effect of thermal fluctuation appears directly in the  Navier-Stokes equations as an \lq\lq external\rq\rq ~force arising from the fluctuating part of the thermodynamic fluxes. 
  
%%%%%%%%%%%  
Bubble nucleation is investigated in a metastable liquid enclosed in a cubic box with periodic boundary conditions, with fixed volume, total mass 
and energy (NVE). 
The equation of state (EoS) we use, which can be chosen freely among available models, e.g. van der Waals or IAPWS \cite{kestin1984thermophysical} EoS for water, 
corresponds to a Lennard-Jones (LJ) fluid \cite{johnson1993lennard} to allow direct comparison with MD simulations.
Quantities are made dimensionless according to $\rho^*=\rho/\rho_r$, $\theta^*=\theta/\theta_r$, $\mathbf{u}^*=\mathbf{u}/U_r$, 
by introducing as reference quantities the parameters of the LJ potential, $\sigma=3.4\times10^{-10}\,{\rm m}$ as length, $\epsilon=1.65\times10^{-21}\,{\rm J} $ as energy,  $m=6.63\times10^{-26}\,{\rm kg}$ as mass, $U_r=(\epsilon/m)^{1/2}$ as velocity, $T_r=\sigma/U_r $ as time, $\theta _r=\epsilon/k_B $ as temperature,  $\mu_r=\sqrt{m\epsilon}/\sigma^2$ as shear viscosity, $c_{vr}=mk_B$ as specific heat at constant volume and $k_r=\mu_r c_{vr}$ as thermal conductivity.  The only dimensionless control group is the capillary number $C=\lambda \rho_r/\sigma^2 U_r^2$, fixed here to $C=5.244$ in order to reproduce the LJ surface tension \cite{johnson1993lennard}.
The symbol $^*$ will be omitted in the following to simplify notation.

\begin{table} [b!]
    \begin{tabular}{  | c | c | c | c | c | c | c l c | c | c | c |}
    \hline
    $\theta_0 $  & $ \rho_L $  &  $R_c $ & $R_c ^{CNT} $ & $ \widetilde {\Delta \Omega}/\theta_0 $ & $ \widetilde{\Delta \Omega}^{CNT}/\theta_0 $ \\ \hline
    $1.25$ & $ 0.45 $& $ 12.8 $ & $ 8.07 $ & $ 1.68 $ & $ 12.89 $   \\ \hline
    $1.25$ & $ 0.46 $& $ 10.2 $ & $ 8.42 $ & $ 5.90 $ & $ 14.05 $ \\ \hline
    $1.25$ & $ 0.47 $& $ 9.5 $   & $ 9.17 $ & $ 13.90 $ & $ 16.67 $ \\ \hline
    $1.25$ & $ 0.48 $& $ 10.1 $  & $ 10.64$ & $ 29.22 $ & $ 22.41 $ \\ \hline
    $1.20$ & $ 0.51 $& $ 7.1$   & $ 6.35$ & $ 10.55 $ & $ 18.13 $ \\ \hline
    $1.20$ & $ 0.52 $& $ 7.0 $   & $ 6.93 $ & $ 20.60 $ & $ 21.60 $ \\ \hline
    \end{tabular}
        \caption{Comparison between CNT and the string method applied to the Diffuse Interface model. Critical radii and (Landau) free energy barriers $\widetilde{\Delta \Omega}$ for bubble nucleation from the liquid. The discrepancy close to the spinodal is well known from the literature.
        \label{table_radii}}
\end{table}
The system volume $V=(600)^3$ has been discretized on an equi-spaced grid with $50$ cells per direction. 
Critical nucleus size, interface thickness and transition time are crucial to prepare a well resolved simulation. 
The preliminary set-up was based on the expected transition time derived from the free energy barriers and on the size $R_c$ of the critical nucleus at the different thermodynamic conditions, as estimated from the string method \cite{weinan2002string} applied to the present system. Technical details are available in the SM \cite{suppl_mat}. The comparison with
the predictions of CNT are summarized in  Tab.~\ref{table_radii}. 
It turns out that, for the present relatively large system, there is no quantitative difference between the barriers estimated from CNT in the different ensembles, and we may use the grand canonical expression $\widetilde{\Delta \Omega}^{CNT}= 4/3 \pi \gamma R_c^2$.
After a convergence analysis we found that a grid size $\Delta x=12$ is sufficient for a reliable simulation in these thermodynamic conditions, see SM \cite{suppl_mat}. Thanks to the extension of the simulated domain, ten runs for each condition provide a reasonably well converged statistics. 

\begin{figure}[t!]
%	\centering
	{\includegraphics[width=0.48\textwidth]{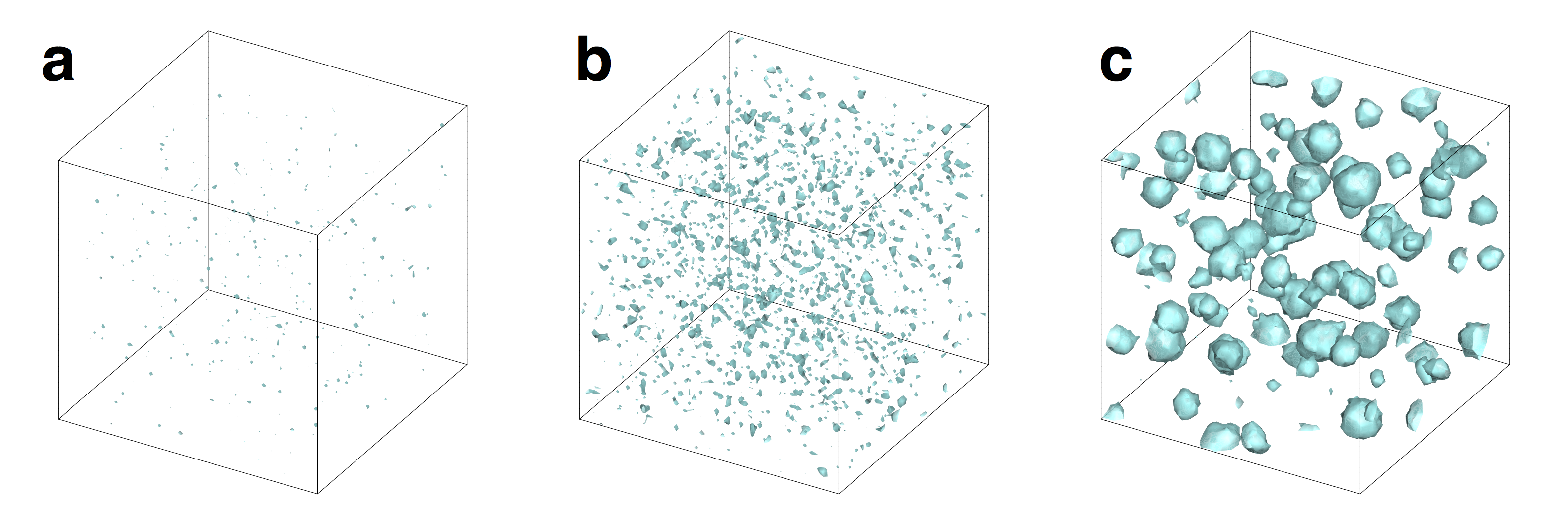}}\\
	{\includegraphics[width=0.48\textwidth]{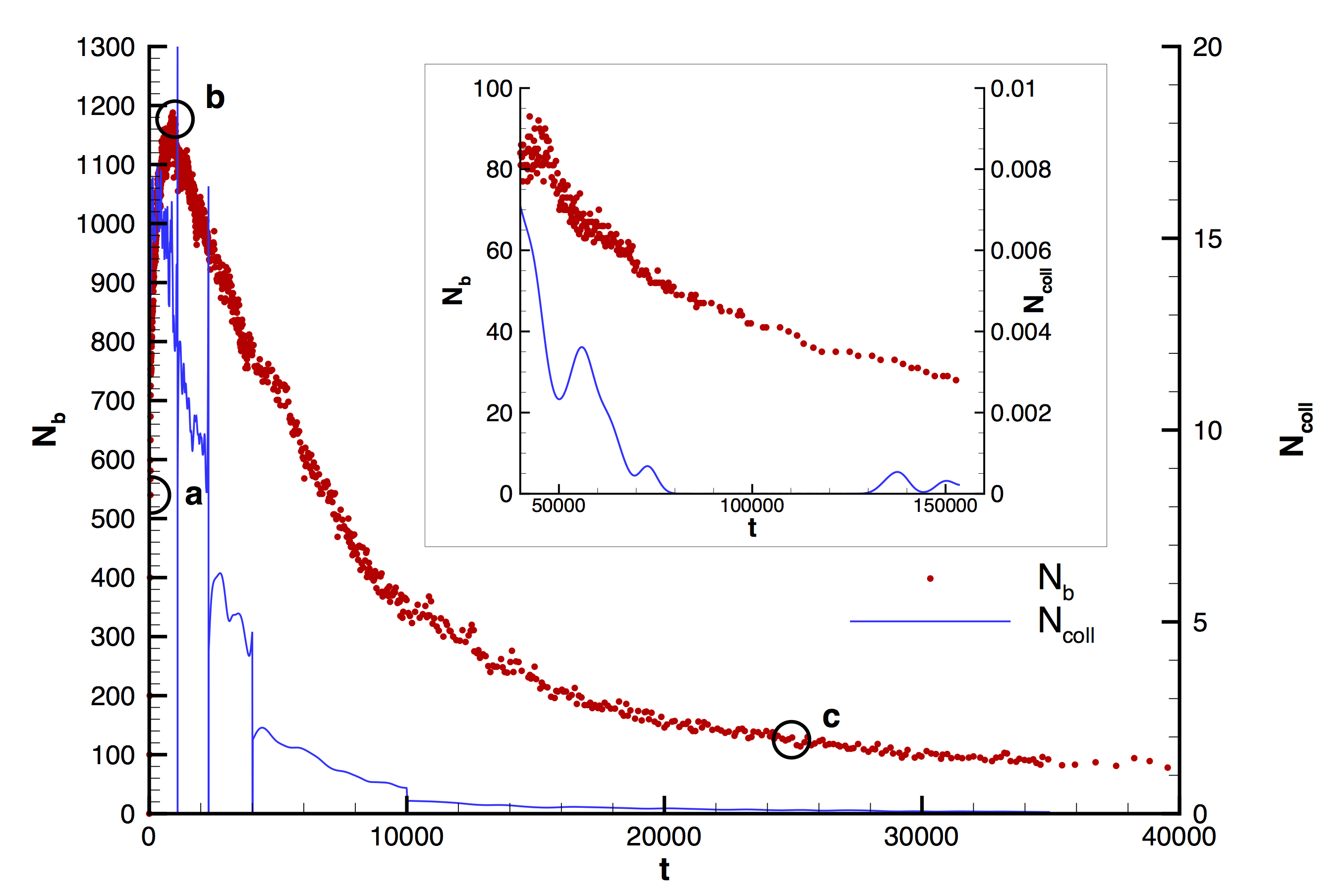}}

	\caption{Top panel: bubble configurations along nucleation ($\rho=0.46,  \; \theta_0=1.25$), from left to right $t=400$, $t=2000$, $t=25000$. Animation available in \cite{suppl_mat}. Bottom panel: bubble number evolution (red symbols) and number of coalescence events (blue line).
		\label{f:nucl_snap_46}}
\end{figure}
\begin{figure}[t!]
%	\centering
	{\includegraphics[width=0.48\textwidth]{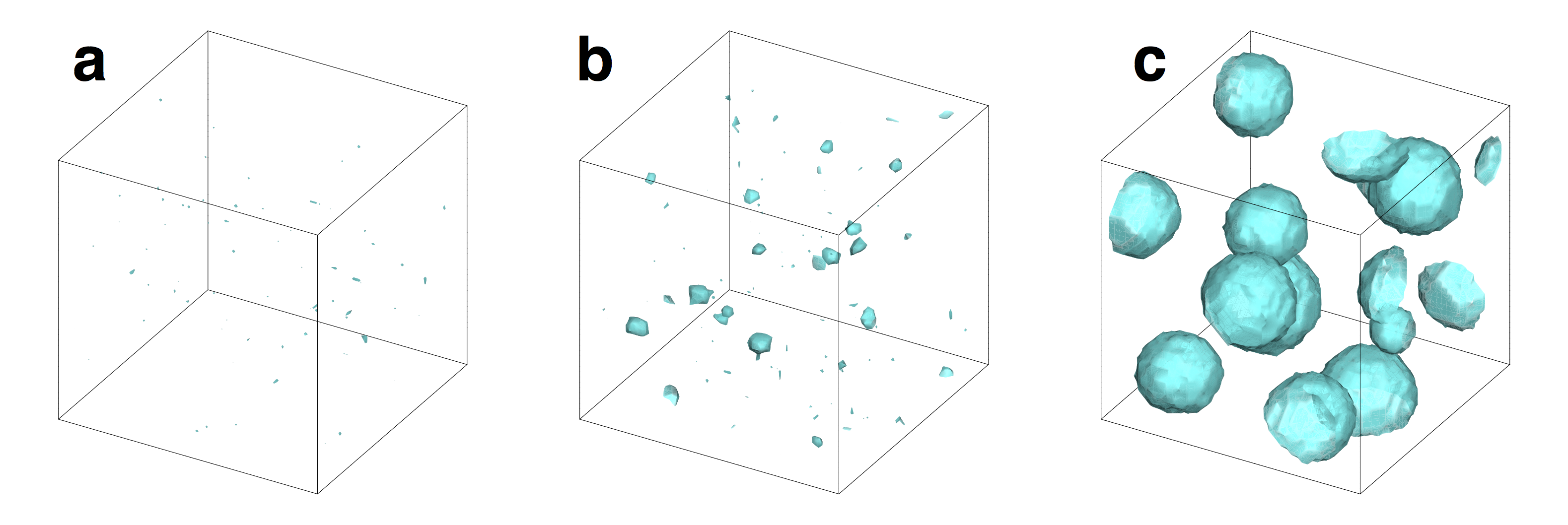}}\\
	{\includegraphics[width=0.48\textwidth]{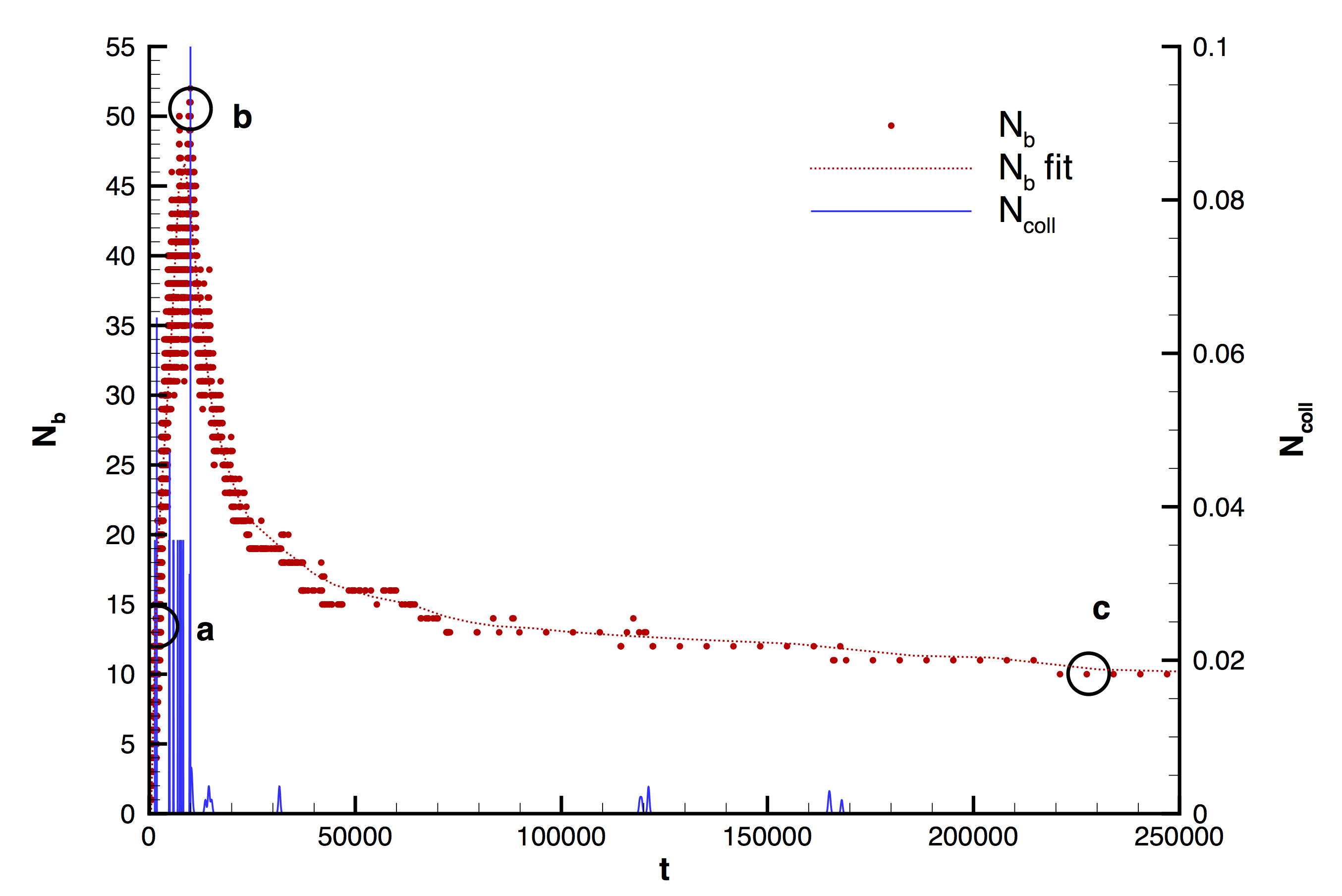}}
	\caption{Same as Figure~\ref{f:nucl_snap_46} at  $\rho=0.48, \; \theta_0=1.25$. Snapshots taken at $t=2000$, $t=11000$, $t=230000$. The bubble number vs time in the bottom panel is fitted by the dotted red line for better readability. 
	\label{f:nucl_rate_48}}
\end{figure}

Among the different conditions we have investigated, we mainly focused on 
the initial temperature $\theta_0=1.25$ at changing bulk density to explore the corresponding metastable range $\rho_L\in\left[\rho_{spin}, \rho_{sat}\right]=\left[0.44, 0.51\right]$, where $\rho_{sat}$ and $\rho_{spin}$ are the saturation and spinodal densities, respectively. 
A few snapshots of the evolution for two different initial conditions are shown in the top panels of Fig.s~\ref{f:nucl_snap_46} and \ref{f:nucl_rate_48}. Starting from a homogeneous metastable liquid phase, the fluctuations lead the system to spontaneously nucleate vapor bubbles.
The nucleii start out with a complex, far from spherical, shape, see, e.g., \cite{diemand2014direct}. 
Roughly,  when they happen to reach a size larger than critical they typically expand. 
Eventually, after a long and complex dynamics where bubbles expand and coalesce, stable equilibrium conditions are reached. 
This new configuration is characterized by several stable vapor bubbles in equilibrium with the surrounding liquid.
The case at $\rho_L = 0.46$, the closest one to the spinodal we considered here, is the more populated, Fig.~\ref{f:nucl_snap_46} in comparison with Fig.~\ref{f:nucl_rate_48}. This system has a lower barrier, hence it nucleates faster. The initial (metastable) thermodynamic condition also influences the number and typical dimension of the bubbles in the final stage, bottom panels of Fig.s~\ref{f:nucl_snap_46} and \ref{f:nucl_rate_48} providing the bubble number $N_b$ as a function of time. 
A tracking procedure has been put forward to follow the evolution of the distinct bubbles. By monitoring volume, mass, center of mass and its velocity, the tracking algorithm allows to detect coalescence events (see \cite{suppl_mat} for the procedure details). The actual number of collisions $\tilde{N}_{coll}$ evaluated at each time step is characterized by a highly discontinuous fingerprint, hence we smoothed it out with a Gaussian kernel with standard deviation on the order of 50 time units to gain more robust indications. The smoothed number of collisions $N_{coll}$, plotted with the blue line in the bottom panel of the Fig.s~\ref{f:nucl_snap_46} and \ref{f:nucl_rate_48}, shows a strong correlation with the number of bubbles throughout the initial \emph{nucleating} stage -- when $N_b$ grows linearly with time (i.e. at a constant nucleation rate) -- and the first part of the expansion phase -- characterized by a rapidly decreasing number of bubbles mainly due to collapse -- that we will call \emph{collapsing} stage. These two stages are characterized by a competing-growth mechanism \cite{tsuda2008} due to the constraint of constant mass, explaining the high number of supercritical bubble collapses. The coalescence events start being less and less probable during the \emph{slowly-expanding} regime that characterizes the long-time dynamics of the multi-bubble system. The inset of the Fig.~\ref{f:nucl_snap_46} zooms into this regime showing that isolated collision events are still occurring, contributing to important acceleration toward the final equilibrium condition. 

\begin{figure}[t!]
%	\centering
	{\includegraphics[width=0.48\textwidth]{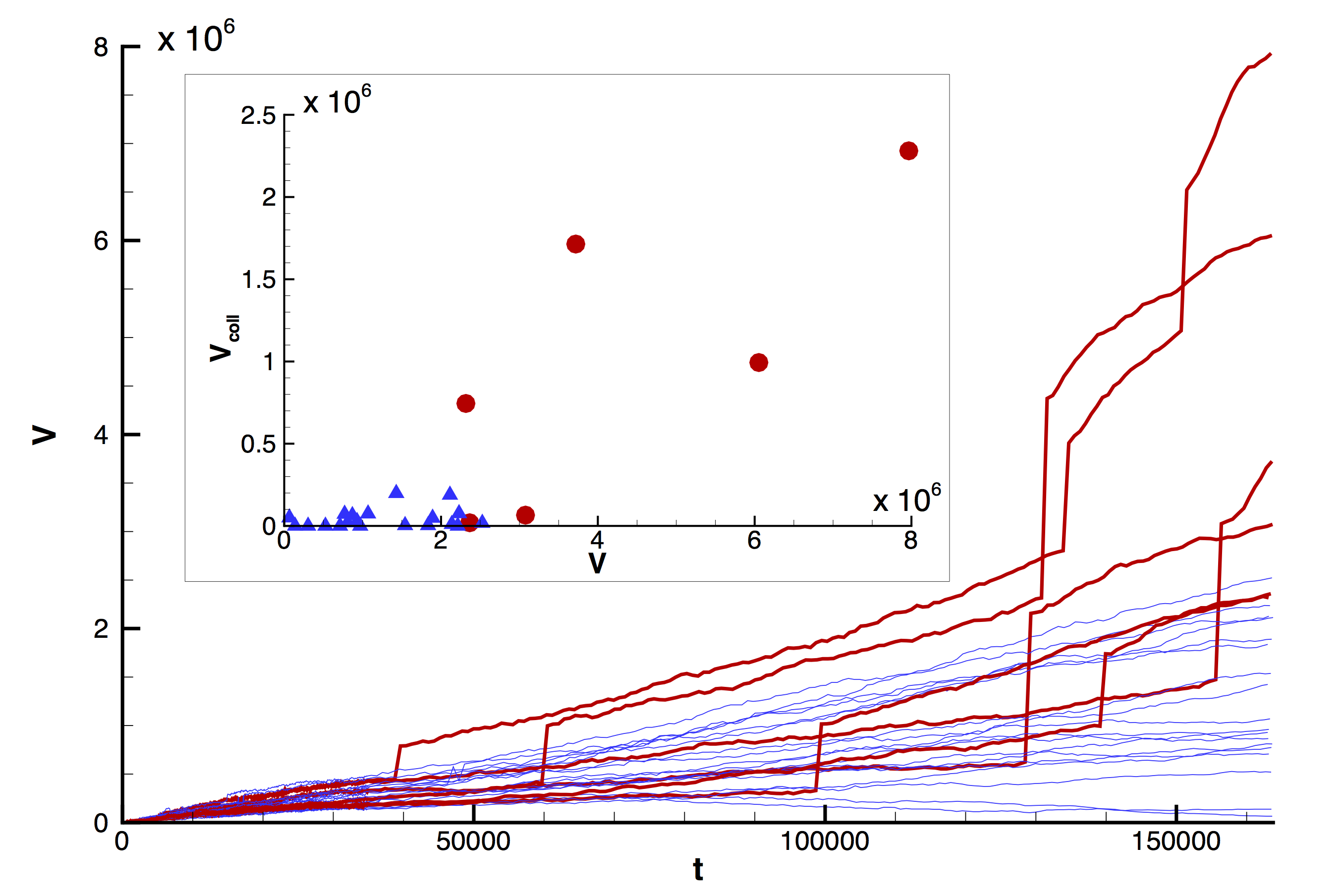}}\\
	\caption{Volume history of the bubbles surviving the entire simulation ($\rho=0.46$ and $\theta_0=1.25$). Intense coalescence events, characterized by a sudden volume jump, are identified in the red curves. The corresponding volumes are shown by the red dots in the inset providing the $V - V_{coll}$ scatter plot, where $V_{coll}$ is the volume acquired by coalescence. The largest bubbles experienced intense coalescence events. 
		\label{f:v-vcoll}}
\end{figure}

The volume history of the distinct bubbles (in particular those survived up to the last time investigated) have been plotted in Fig.~\ref{f:v-vcoll}. Among the different bubble evolutions, we highlighted in red the volume histories of those bubbles that experienced intense coalescence events, characterized by a sudden increase in volume. It is apparent that the larger bubbles gained substantial part of their volume by coalescence.
To substantiate this impression, for each bubble in the last configuration, the sum of the volumes acquired by coalescence throughout the whole evolution, $V_{coll}$, was calculated,  inset of Fig.~\ref{f:v-vcoll}. 

\begin{figure}[t!]
%	\centering
	{\includegraphics[width=0.48\textwidth]{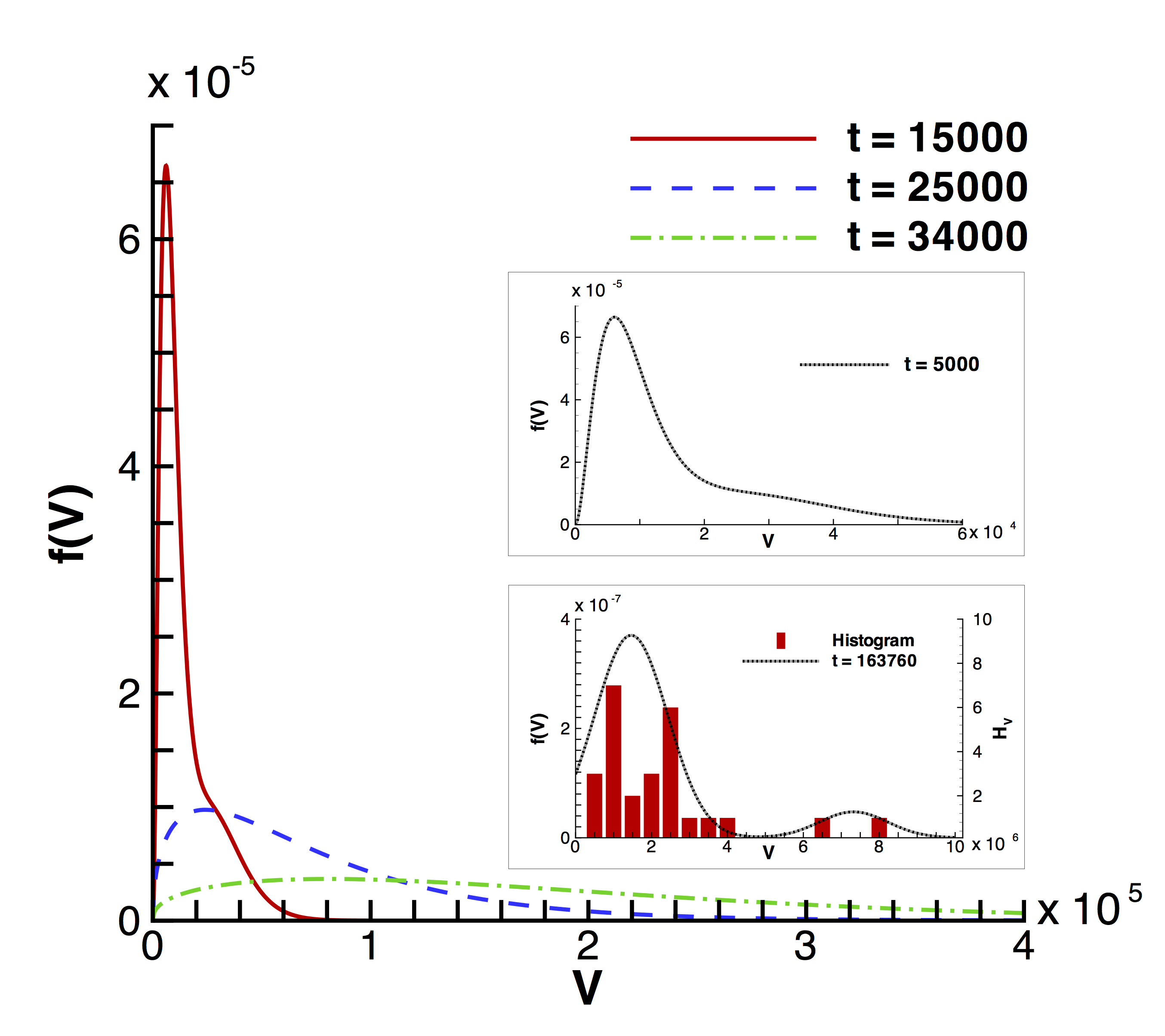}}
	\caption{Probability distribution function $f(V)$ of the bubble volumes during the nucleation, at different times ( 
	$\theta_0 = 1.25$, $\rho = 0.46$, critical volume $V_c = 4445$).
	 		\label{f:pdf}}
\end{figure}
The present mesoscale approach allows to  access the statistics of bubble dimensions.
The probability distribution function of bubble volumes $f(V)$ is plotted in Fig.~\ref{f:pdf}. During both the \emph{nucleating} and \emph{collapsing} stages the pdf is sharply peaked at small volumes, of the order of 2--4 $V_c$. The successive bubble expansion phase is substantially slower and calls for a much longer observation time to detect a significant growth (green dash-dotted curve at $t=34000$). The intense coalescence events explain the presence of the second peak in the pdf at very large volume (black curve in the inset of Fig.~\ref{f:pdf} at $t=163760$).

\begin{figure}[t!]
%	\centering
	{\includegraphics[width=0.48\textwidth]{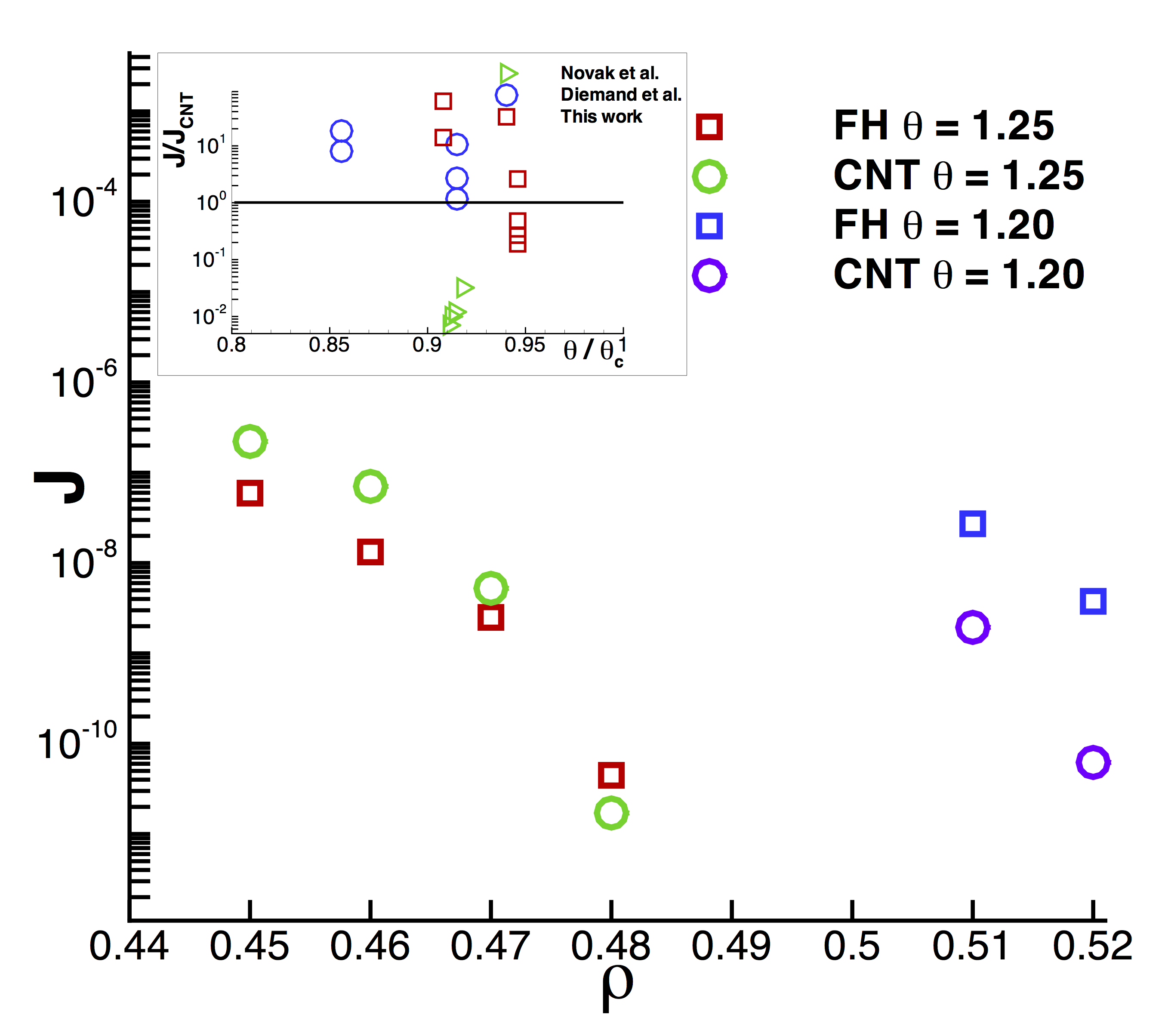}}
	\caption{Nucleation rates: FH simulations (red squares at $\theta_0 =1.25$ and blue squares at $\theta_0 =1.20$) vs CNT (green circles at $\theta_0 =1.25$ and purple circles at $\theta_0 =1.20$). The inset shows the comparison with other authors. 
		\label{f:comparison}}
\end{figure}

The initial \emph{nucleating} stage, where the bubble number increases linearly, gives access to the nucleation rate $J$ in terms of 
bubbles formed per unit time and volume. 
It is here calculated as the slope of the linear fit to the curves of Fig.s~\ref{f:nucl_snap_46} and \ref{f:nucl_rate_48} near the origin.
The CNT expression for the (dimensional) nucleation rate is $J_{CNT} =n_L\sqrt{2\gamma/m\pi}\exp(-\widetilde{\Delta \Omega}/k_B\theta)$, where $n_L$ is the  liquid number density.  It is compared with the measured one in Fig.~\ref{f:comparison} which also provides
the comparison with  some MD simulations \cite{diemand2014direct, novakMD2007}. 
Our results are in reasonable agreement with molecular dynamic simulations in the NVE ensemble \cite{diemand2014direct}, as shown in the inset of Fig.~\ref{f:comparison}, and consistent with the order of magnitude predicted by CNT. The apparent discrepancy that, contrary to expectation, the nucleation rate is smaller where the barriers are smaller than CNT, may be explained by the compression induced by bubble nucleation in the NVE ensemble. \\
In conclusion, the FH approach together with a diffuse interface modeling of the multiphase system have been exploited 
to study homogeneous nucleation of vapor bubbles in metastable liquids. 
We evaluated the nucleation rate and compared it favorably with state of the art simulations and theories. The present technique has revealed extremely cheaper with respect to MD simulations, allowing the analysis of the very long bubble expansion stage where bubble-bubble interaction/coalescence events turn out to determine the eventual bubble size distribution.
The accurate results and the efficiency of the modeling encourage the exploitation to more complex conditions, like e.g. heterogeneous nucleation and multi-species systems, and could pave the way for the development of innovative continuum formulation to address thermally activated processes.

\begin{acknowledgments}
The research leading to these results has received funding from the European Research Council under the European Union's Seventh Framework Programme (FP7/ 2007-2013)/ERC Grant agreement no. [339446].
\end{acknowledgments}
 
\bibliography{mybibfile}

 \end{document}